\documentclass[11pt,floatfix,noshowpacs]{revtex4}
\usepackage[dvips]{graphicx,color}

\usepackage[latin1]{inputenc}
\usepackage{amsmath,amssymb,amsfonts,amsthm,latexsym}
\usepackage{mathrsfs} 
\usepackage[english]{babel}
\usepackage{dsfont}
\usepackage{float}
\usepackage{subfig}
\usepackage{slashed}

\begin{document}

\author{M.F. Izzo Villafañe$^{a,b}$, D. G\'omez Dumm$^{a,b}$ and N.N.\ Scoccola$^{b,c,d}$}

\address{$^{a}$ IFLP, CONICET $-$ Departamento de F\'{\i}sica, Fac.\ de Ciencias
Exactas, Universidad
Nacional de La Plata, C.C. 67, 1900 La Plata, Argentina,\\
$^{b}$ CONICET, Godoy Cruz 2290, 1425 Buenos Aires, Argentina \\
$^{c}$ Physics Department, Comisi\'on Nacional de Energ\'{\i}a Atómica,
Avenida del Libertador 8250, 1429 Buenos Aires, Argentina \\
$^{d}$ Universidad Favaloro, Sol{\'\i}s 453, 1078 Buenos Aires, Argentina}

\title{\sc\Large{Vector and axial vector mesons \\ in a nonlocal chiral quark model}}

\begin{abstract}
Basic features of nonstrange vector and axial vector mesons are
analyzed in the framework of a chiral quark model that includes
nonlocal four-fermion couplings. Unknown model parameters are
determined from some input values of masses and decay constants,
while nonlocal form factors are taken from a fit to lattice QCD
results for effective quark propagators. Numerical results show a
good agreement with the observed meson phenomenology.
\end{abstract}

\maketitle

\section{\sc Introduction}

Given the nonperturbative character of quantum chromodynamics (QCD) in the
low-energy regime, the analysis of hadron phenomenology starting from first
principles is still a challenge for theoretical physics. Although
substantial progress has been achieved in this sense through lattice QCD
(LQCD) calculations, this approach shows significant difficulties, e.g.~when
dealing with small quark masses or with hadronic systems at nonzero chemical
potentials. Thus it is important to study the consistency between the
results obtained through lattice calculations and those arising from
effective models for strongly interacting particles. For two light flavors
it is believed that QCD supports an approximate SU(2) chiral symmetry that
is dynamically broken at low energies, where pions play the role of the
corresponding Goldstone bosons. The well-known Nambu$-$Jona-Lasinio (NJL)
model~\cite{njl,njlrev}, in which light mesons are described as
fermion-antifermion composite states, is a simple effective approach that
shows these features. In the NJL model quarks interact through a local
four-fermion coupling, leading to relatively simple Schwinger-Dyson and
Bethe-Salpeter equations. Now, as a step toward a more realistic approach to
low-energy QCD, it is worth it to consider extensions of the NJL model that
include nonlocal interactions~\cite{ripka}. In particular, this is supported
by lattice calculations, which lead to a given momentum dependence of both
the mass and the wave function renormalization (WFR) in the effective quark
propagators~\cite{Parappilly:2005ei,Furui:2006ks}. It is also seen that
nonlocal extensions of the NJL model do not exhibit some problems that are
present in the local theory. For example, nonlocal interactions regularize
the model in such a way that the effective interaction is finite to all
orders in the loop expansion, thus model predictions are less dependent on
the parameterizations, and there is no need to introduce extra
cutoffs~\cite{Blaschke:1995gr}.

Previous works on nonlocal NJL-like (nlNJL) models, focused on different
aspects of strong interaction physics, can be found in the literature. These
include the study of vacuum hadronic properties considering either
two~\cite{Schmidt:1994di,BB95,Plant:1997jr,Golli:1998rf,Noguera:2005ej,
GomezDumm:2006vz,Noguera:2008,Costa:2010pp} or
three~\cite{Scarpettini:2003fj} active quark flavors, and various nonlocal
form factor shapes. In addition, this framework has been used to describe
the chiral restoration transition for hadronic systems at finite temperature
and/or chemical potential (see e.g.\ Refs.~\cite{GomezDumm:2001fz,
Hell:2008cc,Radzhabov:2010dd,Contrera:2010kz,Hell:2011ic,Carlomagno:2013ona}).
In this work, following the proposal in
Refs.~\cite{Noguera:2005ej,Noguera:2008}, we consider a model in which
nonlocal form factors lead to a momentum dependence of the mass and WFR in
the quark propagator, hence the actual shape of these form factors can be
taken from the data obtained through lattice
calculations~\cite{Noguera:2008,Contrera:2010kz}. We concentrate here in
particular in the incorporation of explicit vector and axial vector
interactions. Therefore, besides the previously considered couplings between
scalar and pseudoscalar quark-antiquark currents, in our model we include
couplings between vector and axial vector nonlocal currents satisfying
proper QCD symmetry requirements. In fact, nonlocal models including vector
and axial vector currents have been previously considered in
Ref.~\cite{Plant:1997jr}. However, those models do not include a
momentum-dependent WFR of quark propagators, which is required in order to
perform the comparison with lattice QCD results. We dedicate the first part
of the paper to work out the formalism in order to derive analytical
expressions for some basic vector meson properties, such as masses and decay
parameters. Then we present numerical results obtained by taking the
nonlocal form factors from a fit to lattice QCD data. It is seen that, after
fixing unknown coupling constants so as to reproduce some input meson
observables, the model provides an adequate phenomenological description of
the considered vector meson properties.

The article is organized as follows. In Sect.\ 2 we introduce the model and
derive the corresponding gap equations at the mean field level. In Sect.\ 3
we describe the vector meson sector, obtaining analytical results for meson
masses and decay amplitudes. The numerical and phenomenological analyses are
included in Sect.\ 4, while in Sect.\ 5 we present a summary of our work.
Finally, in Appendixes A and B we collect some analytical expressions and
describe the calculation procedure.

\section{\sc Model}

We consider a two-flavor chiral quark model that includes nonlocal vector
and axial vector quark-antiquark currents. Since our aim is to choose form
factors that are in agreement with LQCD calculations, it is convenient to
work in Euclidean space, where nonlocal interactions are well
defined~\cite{ripka}. The corresponding effective action is given by
\begin{eqnarray}
\label{se}
S_E &=& \int d^4x \ \left\{ \bar\psi(x)(-\,i \slashed \partial +
\hat{m})\psi(x)-\frac{G_S}{2}\Big[
j_S(x)j_S(x) + \vec j_P(x)\cdot\vec j_P(x) + j_M(x)j_M(x)\Big] \right.
\nonumber\\
    &&\left.
    -\,\frac{G_V}{2}\Big[
\vec j_V^{\,\mu}(x)\cdot\vec j_{V\mu}(x) +
\vec j_A^{\,\mu}(x)\cdot\vec j_{A\,\mu}(x)\Big]
 -\frac{G_0}{2}\; j^{\,0\,\mu}_V(x)j^{\,0}_{V\mu}(x)
 -\frac{G_5}{2}\; j_{A}^{\,0\,\mu}(x)j^{\,0}_{A\, \mu}(x)\right\}\ ,
\end{eqnarray}
where $\psi(x)$ is the $N_f=2$ quark doublet, $\psi = (u\ d)^T$, and $\hat
m={\rm diag}(m_u,m_d)$ is the current quark mass matrix. We will work in the
isospin symmetry limit, assuming $m_u=m_d$, which will be called from now on
$m_c$. The fermion currents are given by~\cite{Noguera:2008}
\begin{eqnarray}\label{eq:currents}
j_S(x)          &=& \int d^4z\; g(z)\,
\bar\psi\left(x+\frac{z}{2}\right)
\psi\left(x-\frac{z}{2}\right)
\ , \nonumber\\
j_P^{a}(x)          &=& \int d^4z\; g(z)\,
\bar\psi\left(x+\frac{z}{2}\right) i\, \gamma_5\, \tau^a
\psi\left(x-\frac{z}{2}\right)
\ , \nonumber\\
j_M(x)          &=& \frac{1}{2\varkappa}\;\int d^4z\; f(z)\,
\bar\psi\left(x+\frac{z}{2}\right)\,
i \overleftrightarrow{\slashed \partial}
\psi\left(x-\frac{z}{2}\right)\ , \nonumber\\
j_{V \mu}^{a}(x) &=& \int d^4z\; h(z)\,
\bar\psi\left(x+\frac{z}{2}\right)\tau^a \gamma_\mu
\psi\left(x-\frac{z}{2}\right)
\ , \nonumber\\
j_{A\, \mu}^{a}(x) &=& \int d^4z\; h(z)\,
\bar\psi\left(x+\frac{z}{2}\right)\tau^a \gamma_\mu \gamma_5
\psi\left(x-\frac{z}{2}\right)
\ , \nonumber\\
j^{\,0}_{V\mu}(x)    &=& \int d^4z\; h_0(z)\,
\bar\psi\left(x+\frac{z}{2}\right) \gamma_\mu
\psi\left(x-\frac{z}{2}\right)
\ , \nonumber\\
j^{\,0}_{A\,\mu}(x)    &=& \int d^4z\; h_5(z)\,
\bar\psi\left(x+\frac{z}{2}\right) \gamma_\mu\gamma_5
\psi\left(x-\frac{z}{2}\right)\ ,
\label{currents}
\end{eqnarray}
where $\tau^a$, $a=1,2,3$, are the Pauli matrices, while
$u(x^\prime)\overleftrightarrow{\partial}v(x)\equiv u(x^\prime)\partial_x
v(x)-\partial_{x^\prime} u(x^\prime)v(x)$. Eqs.~(\ref{currents}) include the
usual scalar ($I=0$) and pseudoscalar ($I=1$) quark-antiquark
currents~\cite{Noguera:2005ej,GomezDumm:2006vz}, as well as vector and
axial-vector quark-antiquark currents that transform as either isospin
singlets or triplets. In addition, we consider a coupling between
``momentum'' currents $j_M(x)$~\cite{Noguera:2005ej,Noguera:2008}, which
involve derivatives of the fermion fields. The presence of this interaction
is naturally expected as a correction arising from the underlying QCD
dynamics. Whereas in a local theory, at the mean field level, it would
simply lead to a redefinition of fermion fields, in our nonlocal scheme it
leads to a momentum-dependent wave function renormalization of the quark
propagator, in consistency with LQCD analyses. For convenience, we have
chosen to take a common coupling constant $G_S$ for both the
scalar/pseudoscalar and momentum quark interaction terms. Notice, however,
that the relative strength between these terms is controlled by the mass
parameter $\varkappa$ in $j_M(x)$. Finally, the functions $f(z)$, $g(z)$,
$h(z)$, $h_0(z)$ and $h_5(z)$ are covariant form factors responsible for the
nonlocal character of the interactions. Notice that, in order to guarantee
chiral invariance, the form factor $g(z)$ has to be equal for the scalar and
pseudoscalar currents $j_S(x)$ and $j^a_P(x)$, and the same applies to the
form factor $h(z)$ entering the vector and axial vector currents
$j_{V\mu}^{a}(x)$ and $j_{A\mu}^{a}(x)$.

To work with mesonic degrees of freedom, we proceed to perform a
bosonization of the fermionic theory~\cite{ripka}. This is done in a
standard way by considering the corresponding partition function
$\mathcal{Z} = \int \mathcal{D}\, \bar{\psi}\mathcal{D}\psi \,\exp[-S_E]$,
and introducing auxiliary bosonic fields $\sigma_1(x)$, $\sigma_2(x)$
[scalar, related respectively to the currents $j_S(x)$ and $j_M(x)$],
$\pi^a(x)$ (pseudoscalar), $v^0_{\mu}(x)$, $v^a_\mu(x)$ (vector) and
$a^0_{\mu}(x)$, $a^a_\mu (x)$ (axial vector), where indices $a$ run from 1
to 3. After integrating out the fermion fields the partition function can be
written as
\begin{eqnarray}
\label{zbos}
\mathcal{Z} &=& \int \mathcal{D}\sigma_1\, \mathcal{D}\sigma_2\,
\mathcal{D}\,\vec{\pi}\, \mathcal{D}\,v^0_{\mu}\, \mathcal{D}\,a^0_{\mu}\,
\mathcal{D}\, \vec{v}_{\mu}\, \mathcal{D}\,\vec{a}_{\mu} \;
\exp\left[-S_E^{\rm bos}\right]
\ ,
\end{eqnarray}
where $S_E^{\rm bos}$ stands for the Euclidean bosonized action.
In momentum space, the latter is given by
\begin{eqnarray}
S^{\rm bos}_{E} &=& -\log\,\det A(p,p^\prime)+\int \dfrac{d^{4}p}{(2\pi)^{4}}
\left\lbrace\dfrac{1}{2G_S}\left[\sigma_1(p)\sigma_1(-p)+
\vec{\pi}(p)\cdot\vec{\pi}(-p)+\sigma_2(p)\sigma_2(-p)\right]\right. \nonumber\\
 & &  \hspace{-0.5cm}
+\left. \dfrac{1}{2G_V} \left[ \vec{v}_{\mu}(p)\cdot\vec{v}^{\,\mu}(-p)+
\vec{a}_{\mu}(p)\cdot\vec{a}^{\,\mu}(-p)\right]
+ \dfrac{1}{2G_0}\;v^0_{\mu}(p) v^{0\mu}(-p)
+ \dfrac{1}{2G_5}\;a^0_{\mu}(p) a^{0\mu}(-p)
\right\rbrace ,
\label{sboson}
\end{eqnarray}
where the operator $A(p,p^\prime)$ reads
\begin{eqnarray}
\label{A}
A(p,p^\prime)&=& (2\pi)^4 \delta^{(4)}(p-p^\prime)(-\slashed p+m_c) +
                 g(\bar p)\, \bigg[\sigma_1(p^\prime-p)+i \gamma_5
                 \vec{\tau} \cdot \vec{\pi}(p^\prime-p)\bigg]
                 \nonumber\\
             & &
             +\ f(\bar p)\,\dfrac{\slashed {\bar p}}{\varkappa}\,\sigma_2(p^\prime-p)\
             +\ h(\bar p)\, \gamma^{\mu}\bigg[
             \vec{\tau} \cdot \vec{v}_{\mu}(p^\prime-p) +
             \gamma_5\, \vec{\tau} \cdot \vec{a}_{\mu}(p^\prime-p)\bigg]
                 \nonumber\\
             & &
             +\ h_0(\bar p)\,\gamma^{\mu}\; v^0_{\mu}(p^\prime-p)
             +\ h_5(\bar p)\,\gamma^{\mu}\gamma_5\; a^0_{\mu}(p^\prime-p)
             \ ,
\end{eqnarray}
with $\bar p \equiv (p+p')/2$. Here, the functions $f(p)$, $g(p)$,
$h(p)$, $h_0(p)$, and $h_5(p)$ stand for the Fourier transforms of
the form factors entering the nonlocal currents in
Eq.~(\ref{currents}). Without loss of generality, the coupling
constants can be chosen so that the form factors are normalized to
$f(0) = g(0) = h(0) = h_0(0) = h_5(0) = 1$.

Let us now consider the mean field approximation (MFA), in which the bosonic
fields are expanded around their vacuum expectation values, $\phi(x) = \bar
\phi + \delta\phi(x)$. On the basis of charge, parity and Lorentz
symmetries, we assume that $\sigma_1(x)$ and $\sigma_2(x)$ have nontrivial
translational invariant mean field values $\bar{\sigma}_1$ and
$\varkappa\,\bar\sigma_2$, respectively, while the vacuum expectation values
of the remaining bosonic fields are zero (notice that $\bar\sigma_2$ is
dimensionless, due to the introduction of the parameter $\varkappa$).
Writing the operator $A(p,p')$ as $A = A_0 + \delta A$, within this
approximation one can expand the logarithm of the fermionic determinant as
\begin{equation}
\log\det A \ = \ {\rm tr}\,\log A \ = \ {\rm tr}\,\log A_0 + {\rm tr}\,
(A_0^{-1}\delta A) - \,\frac{1}{2}\,{\rm tr}\,(A_0^{-1}\delta A\,
A_0^{-1}\delta A) + \dots \ ,
\end{equation}
where
\begin{equation}
A_0(p,p') \ = \ (2\pi)^4 \delta^{(4)}(p-p^\prime)\left\{-[1-\bar\sigma_2\, f(p)]\,
\slashed p + m_c + \bar\sigma_1\, g(p)\right\}\ ,
\label{acero}
\end{equation}
and the trace extends over Dirac, color, flavor and momentum spaces. In the
same way, the bosonized effective action in Eq.~(\ref{sboson}) can be
expanded in powers of meson fluctuations as
\begin{equation}\label{expansion}
S_{E}^{\rm bos}\ =\ S_{E}^{\,\mbox{\tiny{\rm MFA}}}\;+\;S_{E}^{\rm
quad}\;+\;\dots\ ,
\end{equation}
where the mean field action per unit volume reads~\cite{Noguera:2008}
\begin{equation}
\dfrac{S_E^{\,\mbox{\tiny{\rm MFA}}}}{V^{(4)}} \ = \ -2\,N_C \int
\dfrac{d^4 p}{(2\pi)^4}\ \mathrm{Tr} \log
[\mathcal{D}_0^{-1}(p)]\; + \dfrac{1}{2G_S}\big(\bar{\sigma}_1^2
\; + \varkappa^2\bar{\sigma}_2^2\big)\ ,
\label{sr}
\end{equation}
the trace acting just over Dirac space. From Eq.~(\ref{acero}), the mean
field effective quark propagator $\mathcal{D}_0(p)$ is given by
\begin{equation}
\mathcal{D}_0(p)=\dfrac{z(p)}{-\slashed p + m(p)} \ ,
\end{equation}
where the functions $m(p)$ and $z(p)$ ---momentum-dependent
effective mass and WFR--- are related to the nonlocal form factors and
the vacuum expectation values of the scalar fields by
\begin{eqnarray}
z(p) &=& \left[ 1\,-\,\bar \sigma_2 \,f(p)\right]^{-1}\ ,
\nonumber\\
m(p) &=& z(p)\, \left[ m_c\, +\, \bar \sigma_1\, g(p)\right] \ .
\label{mz}
\end{eqnarray}

The mean field values $\bar{\sigma}_{1,2}$ can be found by
minimizing the mean field Euclidean action. This leads to the set of
coupled gap equations~\cite{Noguera:2008}
\begin{eqnarray}
\bar\sigma_1 & = & 8\, N_C\, G_S \int \frac{d^4 p}{(2\pi)^4}\
g(p)\;\frac{z(p)\,m(p)}{D(p)}\ ,\nonumber \\
\bar\sigma_2 & = & -\,8\, N_C\, G_S \int \frac{d^4 p}{(2\pi)^4}\
\dfrac{p^2}{\varkappa^2}\;f(p)\;\frac{z(p)}{D(p)}\ ,
\label{gapeq}
\end{eqnarray}
where we have defined $D(p) = p^2 + m(p)^2$. The chiral quark condensates
---order parameters of the chiral restoration transition--- are given by the
vacuum expectation values $\langle \bar q q\rangle$, where $q = u,d$. The
corresponding expressions can be obtained by differentiating the MFA
partition function with respect to the current quark masses. Away from the
chiral limit, this leads in general to divergent integrals. Since one is
interested in the description of the nontrivial vacuum properties arising
from strong interactions, it is usual to regularize these integrals by
subtracting the free quark contributions (see e.g.\
Refs.~\cite{Plant:1997jr,Noguera:2005ej,Hell:2008cc,Radzhabov:2010dd}). One
gets in this way
\begin{equation}
\langle \bar q q\rangle \ = \ -\, 4\, N_C \int \frac{d^4p}{(2\pi)^4}
\left(\frac{z(p)\, m(p)}{D(p)}\; -\; \frac{m_c}{p^2+m_c^2}\right)
\ .
\end{equation}

\section{\sc Meson masses and decay constants}

We are interested in the description of vector meson phenomenology, which
requires going beyond the MFA. In this section we derive analytical
expressions to be used for the calculation of basic measurable
phenomenological quantities, such as meson masses and decay constants. It is
important to notice that pion observables, already calculated within this
framework in previous works~\cite{Noguera:2005ej,Noguera:2008,Dumm:2010hh},
need to be revisited owing to the mixing between $\vec\pi$ and $\vec a_\mu$
fields.

\subsection{\sc Meson masses and mixing}

In general, meson masses can be obtained from the terms in the
Euclidean action that are quadratic in the bosonic fields. When
expanding the bosonized action we obtain
\begin{eqnarray} \label{eq:quad}
S_E^{\rm quad} &=& \dfrac{1}{2} \int \frac{d^4 p}{(2\pi)^4}\
\Big\{ G_\sigma(p^2)\, \delta\sigma(p)\, \delta\sigma(-p)  +
G_{\sigma^\prime}(p^2)\, \delta\sigma^\prime(p)\,\delta\sigma^\prime(-p) \nonumber\\
                && +\;G_{\pi}(p^{2}) \, \delta\vec\pi(p)\cdot\delta\vec\pi(-p) +
                 i\,G_{\pi a}(p^{2})\Big[p^\mu\,\delta\vec{a}_{\mu}(-p) \cdot
                \delta\vec{\pi}(p)-p^\mu\,\delta\vec{a}_{\mu}(p)
                \cdot \delta\vec{\pi}(-p)\Big] \nonumber \\
                & & +\; G_0^{\mu\nu}(p^2)\,\delta v^0_{\mu}(p)\,\delta
                v^0_{\nu}(-p) + G_5^{\mu\nu}(p^2)\,\delta a^0_{\mu}(p)\,\delta
                a^0_{\nu}(-p)\nonumber \\
                & & + \;G_v^{\mu\nu}(p^2)\,\delta\vec{v}_\mu(p)\cdot\delta\vec{v}_\nu(-p) +
G_{a}^{\mu\nu}(p^2)\,\delta\vec{a}_\mu(p)\cdot\delta\vec{a}_\nu(-p)\Big\}\ ,
\label{sequad}
\end{eqnarray}
where the functions $G_M(p^2)$, $M = \sigma,\sigma',\pi,\dots$ are given by
one-loop integrals arising from the fermionic determinant in the bosonized
action. In the case of the $\sigma_1,\sigma_2$ sector the expression in
Eq.~(\ref{sequad}) is given in terms of the fields $\sigma$ and
$\sigma^\prime$, which are defined as linear combinations of $\sigma_1$ and
$\sigma_2$,
\begin{equation}
\delta\sigma = \cos \theta \; \delta\sigma_1 - \sin \theta \; \delta\sigma_2 \ ,
\qquad\ \delta\sigma^\prime = \sin \theta' \; \delta\sigma_1 +
\cos \theta' \; \delta\sigma_2 \ .
\end{equation}
The mixing angles $\theta$ and $\theta'$ are fixed in such a way that there
is no $\sigma - \sigma^\prime$ mixing terms at the level of the quadratic
action for $p^2 = -m_{\sigma^{(\prime)}}^2$, where the minus sign is due to
the fact that the action is given in Euclidean space. Once cross terms have
been eliminated, the functions $G_M(p^2)$ stand for the inverses of the
effective meson propagators, thus scalar meson masses are obtained by
solving the equations $G_{\sigma^{(\prime)}}(-m_{\sigma^{(\prime)}}^2) = 0$.
Explicit expressions for the functions $G_{\sigma^{(\prime)}}(p^2)$ can be
found in Ref.~\cite{Noguera:2008}.

To analyze the vector meson sector one has to take into account the tensors
$G_v^{\mu\nu}$, $G_a^{\mu\nu}$, $G_0^{\mu\nu}$ and $G_{5}^{\mu\nu}$. From
the expansion of the fermionic determinant we obtain
\begin{eqnarray}
G_v^{\mu\nu}(p^2) &=& G_{\rho}(p^2)\left(g^{\mu\nu}-\dfrac{p^{\mu}p^{\nu}}{p^2}\right)+
L_{+}(p^2)\dfrac{p^{\mu}p^{\nu}}{p^{2}}\ ,  \nonumber\\
G_{a}^{\mu\nu}(p^2) &=& G_{{\rm a}_1}(p^2)\left(g^{\mu\nu}-\dfrac{p^{\mu}p^{\nu}}{p^{2}}\right)+
L_{-}(p^{2})\dfrac{p^{\mu}p^{\nu}}{p^{2}}\ ,
\end{eqnarray}
where
\begin{eqnarray}
\hspace{-4mm} G_{\rho \choose {\rm a}_1}(p^{2})&=& \dfrac{1}{G_V}-8N_{C}\, \int
                               \dfrac{d^{4}q}{(2\pi)^{4}} \, h^{2}(q)\,
                               \dfrac{z(q^+)z(q^-)}{D(q^{+})D(q^{-})} \,
                               \left[\dfrac{q^{2}}{3}+\dfrac{2(p\cdot
                               q)^{2}}{3p^{2}}-
                               \dfrac{p^{2}}{4}\pm m(q^{-})m(q^{+})\right],
                               \label{grho} \\
\hspace{-4mm} L_{\pm}(p^{2})&=&\dfrac{1}{G_V}-8N_{C} \, \int \dfrac{d^{4}q}{(2\pi)^{4}} \, h^{2}(q)\,
                 \dfrac{z(q^+)z(q^-)}{D(q^{+})D(q^{-})} \,
                 \left[q^{2}-\dfrac{2(p\cdot q)^{2}}{p^{2}}+\dfrac{p^{2}}{4}\pm
                 m(q^{-})m(q^{+})\right] ,
                 \label{lpm}
\end{eqnarray}
with $q^\pm = q \pm p/2$. The functions $G_{\rho,{\rm a}_1}(p^{2})$ and
$L_{\pm}(p^{2})$ correspond to the transverse and longitudinal projections
of the vector and axial vector fields, describing meson states with spin 1
and 0, respectively. Thus the masses of the physical $\rho^0$ and $\rho^\pm$
vector mesons (which are degenerate in the isospin limit) can be obtained by
solving the equation
\begin{equation}
G_\rho(-m_\rho^2)\ =\ 0 \ .
\label{rmass}
\end{equation}
In addition, in order to obtain the physical states, the vector
meson fields have to be normalized through
\begin{equation}
\delta v^a_\mu(p) = Z_\rho^{1/2}\;{\tilde v}^a_\mu(p)\ ,
\end{equation}
where
\begin{equation}
\label{zr}
Z_\rho^{-1}= g_{\rho qq}^{-2} = \frac{dG_\rho(p^2)}{dp^2}\bigg\vert_{p^2=-m_\rho^2} \ .
\end{equation}
Here $g_{\rho qq}$ can be viewed as an effective $\rho$ meson-quark
effective coupling constant. Regarding the isospin zero channels, it is easy
to see that the expressions for $G_0^{\mu\nu}(p^2)$ can be obtained from
those for $G_v^{\mu\nu}(p^2)$, just replacing $G_V\to G_0$ and $h(q)\to
h_0(q)$. In this way, one can define for the $\omega$ vector meson a function
$G_\omega(p^2)$, obtaining the $\omega$ mass and wave function
renormalization as in Eqs.~(\ref{rmass}) and (\ref{zr}). Similar relations
apply to the axial vector sector, where $G_5^{\mu\nu}(p^2)$ can be obtained
from $G_a^{\mu\nu}(p^2)$ by replacing $G_V\to G_5$ and $h(q)\to h_5(q)$. The
lightest physical state associated to this sector (quantum numbers $I=0$,
$J^P=1^+$) is the $f_1$ axial vector meson, hence we denote by
$G_{f_1}(p^2)$ the form factor corresponding to the transverse part of
$G_5^{\mu\nu}(p^2)$.

In the case of the pseudoscalar sector, from Eq.~(\ref{sequad}) it is seen
that there is a mixing between the pion fields and the longitudinal part of
the axial vector fields~\cite{Ebert:1985kz,Bernard:1993rz}. The mixing term
includes a loop function $G_{\pi a}(p^2)$, while the term quadratic in
$\delta\pi$ is proportional to the loop function $G_\pi(p^2)$. These
functions are given by
\begin{eqnarray}
G_{\pi}(p^2) & = & \dfrac{1}{G_S} \, - \, 8N_{C} \int
\dfrac{d^{4}q}{(2\pi)^{4}}\; g(q)^2\,
                 \dfrac{z(q^+)z(q^-)}{D(q^{+})D(q^{-})}\,
                 \left[(q^{+}\cdot q^-)\,+\,m(q^{+})\,m(q^{-})\right] \ , \nonumber \\
G_{\pi a}(p^2) & = & \dfrac{8N_{C}}{p^{2}}\, \int
\dfrac{d^{4}q}{(2\pi)^{4}}\, g(q)\,h(q)\,
                 \dfrac{z(q^+)z(q^-)}{D(q^{+})D(q^{-})}\,
                 \left[(q^{+}\cdot p)\,m(q^{-})-(q^{-}\cdot
                 p)\,m(q^{+})\right] \ ,
\end{eqnarray}
where once again we have used the definitions $q^\pm = q \pm p/2$. The
physical states $\tilde{\vec{a}}_\mu$ and $\tilde{\vec{\pi}}$ can be now
obtained through the relations~\cite{Ebert:1985kz,Bernard:1993rz}
\begin{eqnarray}
\delta\pi^b(p)     &=& Z^{1/2}_\pi\;{\tilde \pi}^b(p) \ ,\nonumber\\
\delta a^b_{\mu}(p) &=& Z^{1/2}_a\;{\tilde a}^b_\mu(p) - i \, \lambda(p^2) \,  p_\mu \,
Z^{1/2}_\pi\;{\tilde \pi}^b(p) \ ,
\label{pia1mixing}
\end{eqnarray}
where the mixing function $\lambda(p^2)$, defined in such a way
that the cross terms in the quadratic expansion vanish, is given
by
\begin{equation}
\lambda(p^2) = \dfrac{G_{\pi a}(p^2)}{L_-(p^2)} \ .
\end{equation}
The pion mass can be then calculated from $G_{\tilde
\pi}(-m_\pi^2) = 0$, where
\begin{equation}
G_{\tilde{\pi}}(p^2)= G_{\pi}(p^2)-\dfrac{G_{\pi
a}^2(p^2)}{L_-(p^2)}\,p^2\ ,
\end{equation}
while the pion WFR can be obtained from
\begin{equation}
\label{zpi} Z_\pi^{-1}= g_{\pi qq}^{-2} = \frac{dG_{\tilde
\pi}(p^2)}{dp^2}\bigg\vert_{p^2=-m_\pi^2} \ .
\end{equation}
In the case of the ${\rm a}_1$ axial vector mesons ($I=1$ triplet), since
the transverse parts of the $a_\mu^b$ fields do not mix with the pions, the
corresponding mass and WFR can be calculated using relations analogous to
those quoted for the vector meson sector, namely Eqs.~(\ref{rmass}) and
(\ref{zr}), with $G_{{\rm a}_1}(p^2)$ given by Eq.~(\ref{grho}).

\subsection{\sc Pion weak decay}

By definition the pion weak decay constant $f_\pi$ is given by the
matrix elements of axial currents between the vacuum and the
physical one-pion states,
\begin{equation}
\label{eq:fpi} \langle 0 \vert {\cal J}_{A\mu}^a (x) \vert
\tilde{\pi}^b(p) \rangle= i \, e^{-i p\cdot x}\,\delta^{ab} \,
f_\pi(p^2) \; p_\mu \ ,
\end{equation}
evaluated at the pion pole. To determine the axial currents, we ``gauge'' the
effective action $S_E$, introducing external gauge fields. In general, for a
local theory, this is carried out just by replacing
\begin{equation}
\partial_\mu \longrightarrow \partial_\mu + i\, {\cal G}_\mu\ ,
\label{cov}
\end{equation}
where ${\cal G_\mu}$ is the corresponding gauge field.
In our model, due to the nonlocality of the interactions, the
gauging procedure requires the introduction of gauge fields not
only through the covariant derivative in Eq.~(\ref{cov}) but also
through a parallel transport of the fermion fields in the nonlocal
currents (see e.g.~Refs.~\cite{ripka,BB95,GomezDumm:2006vz}):
\begin{eqnarray}
\psi(x-z/2) & \to & W_G(x,x-z/2)\;\psi(x-z/2)\ , \nonumber \\
\psi^\dagger(x+z/2) & \to & \psi^\dagger(x+z/2)\;W_G(x+z/2,x)\ .
\label{nltransf}
\end{eqnarray}
Here $x$ and $z$ are the variables in the definitions of the
nonlocal currents in Eq.~(\ref{currents}), while the function
$W_G(x,y)$ is defined by
\begin{equation}
W_G(x,y) \ = \ P\;\exp\left[i \,\int_x^y ds^\mu\; {\mathcal
G}_\mu(s)\right] \ , \label{path}
\end{equation}
where $s$ runs over an arbitrary path connecting $x$ with $y$. In
the case of the axial current we introduce the axial gauge fields
${\mathcal W}_\mu^{\,a}(x)$, taking
\begin{equation}
{\cal G}_\mu \ = \dfrac{1}{2}\ \gamma_5 \
\vec{\tau} \cdot \vec{\mathcal{W}}_\mu \ .
\end{equation}
In addition, notice that if the action is written in terms of the original
states $\pi^b$ and $a_\mu^b$, in order to calculate the matrix element in
Eq.~(\ref{eq:fpi}) one has to take into account the mixing described in the
previous subsection. Once the gauged effective action is built, the matrix
elements can be obtained by taking derivatives with respect to the gauge and
the physical pion fields,
\begin{equation}
\langle 0 \vert {\cal J}_{A\mu}^a(x) \vert \tilde\pi^b(p) \rangle
\ = \ \frac{\delta^2 S_E^{\rm bos}}{\delta\mathcal{W}_\mu^a(x)\, \delta\tilde\pi^b(p)}
\bigg\vert_{\mathcal{W}_\mu^a=\tilde\pi^b=0}\ .
\end{equation}

\begin{figure}[h]
\subfloat{\includegraphics[scale=0.65]{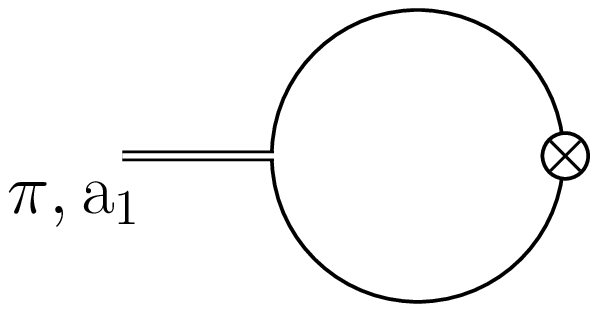}} \hspace{1cm}
\subfloat{\includegraphics[scale=0.65]{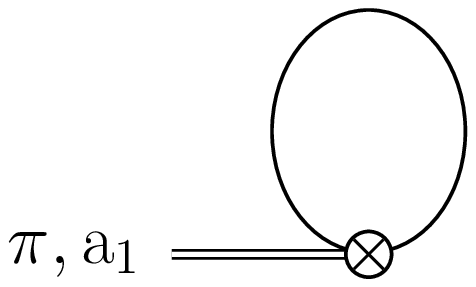}}\\
\caption{Diagrammatic representation of the contributions to the
pion decay constant. The cross represents the axial current
vertex.}
\label{fig:fpigraf}
\end{figure}
The resulting one-loop contributions are diagrammatically
schematized in Fig.~\ref{fig:fpigraf}. Tadpole-like diagrams,
which are not present in the local NJL model, arise from the
occurrence of gauge fields in Eqs.~(\ref{nltransf}). We finally
obtain
\begin{equation}
f_\pi = \dfrac{m_c \, g_{\pi q \bar{q}}}{m_\pi^2} \left[ F_0 (-m_\pi^2)
+ \lambda(p^2) \, F_1 (-m_\pi^2)\right] \ ,
\label{fpi}
\end{equation}
where
\begin{eqnarray}
F_0 (p^2) &=& 8N_c\,\int \dfrac{d^4 q}{(2\pi)^4}\,g(q)\, \dfrac{z(q^+)z(q^-)}{D(q^+)D(q^-)}
              \left[ (q^+\cdot q^-) + m(q^+)\,m(q^-)\right] \ ,\nonumber\\
F_1 (p^2) &=& 8N_c\, \int \dfrac{d^4 q}{(2\pi)^4}\,h(q)\, \dfrac{z(q^+)z(q^-)}{D(q^+)D(q^-)}
              \left[ (q^+\cdot p)\,m(q^-) - (q^-\cdot
              p)\,m(q^+)\right]\ .
\label{f01}
\end{eqnarray}
It is important to notice that the result for $f_\pi$ does not depend on the
path chosen for the transport function in Eq.~(\ref{path}) [see the comment
after Eq.~(\ref{alphalong}) below]. In the absence of vector meson fields, the
mixing term in Eq.~(\ref{fpi}) vanishes and our expression reduces to that
previously quoted in Ref.~\cite{Noguera:2008}.

\subsection{\sc $\rho$ meson-photon vertex and $\rho$ electromagnetic decay constant}

Another important quantity to be studied is the $\rho$-photon vertex. In our
nonlocal model, meson-photon couplings receive in general contributions from
the parallel transport in Eq.~(\ref{nltransf}), therefore we find it important
to check that the conservation of the vector current is satisfied. In
addition, from this vertex we can obtain a prediction for the
electromagnetic $\rho\to e^+e^-$ decay amplitude.

The $\rho$-photon vertex is given by the matrix element of the
electromagnetic current between a vector meson state and the vacuum,
\begin{equation}
\langle 0 \vert {\cal J}_{{\rm em}\,\mu}(x) \vert \tilde v_\nu^a(p)
\rangle \ = \ i\, e^{-ip\cdot x}\, \Pi_{\mu\nu}^a(p) \ .
\end{equation}
To calculate this matrix element one can follow the procedure discussed in
the previous subsection, taking now
\begin{equation}
{\cal G}_\mu \ = \ e\, Q\, {\mathcal A}_\mu \ ,
\end{equation}
where $e$ is the proton charge and $Q = {\rm diag}(2/3\; ,\; -1/3)$.

\begin{figure}[h]
\begin{center}
\subfloat{\includegraphics[scale=0.65]{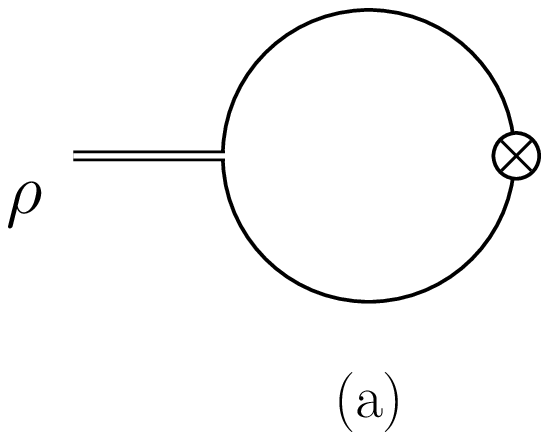}} \hspace{1cm}
\subfloat{\includegraphics[scale=0.65]{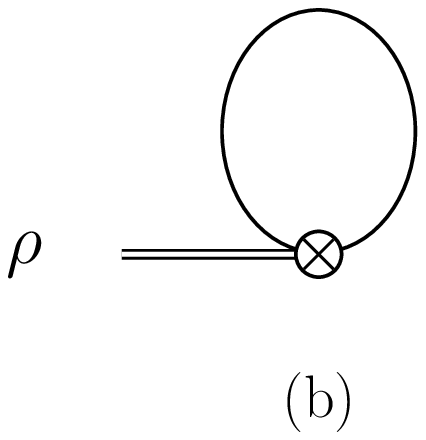}}\\
\end{center}
\caption{Diagrams contributing to the $\rho$ meson-photon vertex.}
\label{fig:ward}
\end{figure}

Once again it is possible to distinguish two contributions to
$\Pi_{\mu\nu}^a$, namely $\Pi_{\mu\nu}^{{\rm (I)}\,a}$ and
$\Pi_{\mu\nu}^{{\rm (II)}\,a}$, arising from a two-vertex and a
tadpole-like diagram, respectively (see Fig.~\ref{fig:ward}). We
obtain
\begin{eqnarray}
\label{eq:wardp1}
\Pi_{\mu\nu}^{{\rm (I)}\,a} (p) & = & 4N_C\, \delta_{a3}\, e\, Z_\rho^{1/2}
     \int \dfrac{d^4 q}{(2\pi)^4}\,
     \dfrac{z(q^+)z(q^-)}{D(q^+)D(q^-)}\; h(q)\;\nonumber \\
     & & \times \Bigg\lbrace \frac{1}{2}\Big[\dfrac{1}{z(q^+)} +
     \dfrac{1}{z(q^-)}\Big] \Big[q^+_\mu\, q^-_\nu + q^+_\nu\, q^-_\mu
     - (q^+ \cdot q^-)\, \delta_{\mu\nu}- m(q^+)m(q^-)\,\delta_{\mu\nu}\Big] \nonumber \\
     & &  + \; \bar\sigma_1 \Big[ m(q^+)\,q^-_\nu + m(q^-)\,q^+_\nu \Big]\; \alpha_{g\,\mu}(q,p) \nonumber \\
     & &  + \; \bar\sigma_2 \Big[- \frac{(q^-)^2}{2}\,q_\nu^+
      - \frac{(q^+)^2}{2}\, q_\nu^- + m(q^+)m(q^-)\,q_\nu\Big]
      \alpha_{f\,\mu}(q,p) \Bigg\rbrace \ , \\
\Pi_{\mu\nu}^{{\rm (II)}\,a} (p) & = & -\,
4N_C\, \delta_{a3} \, e\, Z_\rho^{1/2} \int \dfrac{d^4 q}{(2\pi)^4}\, \dfrac{z(q)}{D(q)}
\ q_\nu \, \alpha_{h\,\mu}(q,p) \ .
\end{eqnarray}
Here we have defined, for a given function $f(p)$,
\begin{equation}
\alpha_{f\,\mu}(q,p) \ = \
\int \dfrac{d^4 \ell}{(2\pi)^4}\; \Big[f(q+\ell/2)\, F_\mu(p-\ell,\ell)
+ f(q-\ell/2)\, F_\mu(\ell,p-\ell)\Big] \ ,
\end{equation}
with
\begin{equation}
F_\mu(k,k') \ = \ -\, i \int d^4z\; e^{i k'z}\int_0^z ds_\mu\;
e^{-i(k+k')s} \ ,
\label{pathf}
\end{equation}
where $s$ runs over a path connecting the origin with a point
located at $z$.

It can be seen that the tensors $\Pi_{\mu\nu}^{{\rm (I)}\,a}$ and
$\Pi_{\mu\nu}^{{\rm (II)}\,a}$ are in general not transverse. However, the
sum of both contributions satisfies $p^\mu \,\Pi_{\mu\nu}^a = 0$, as
required from the conservation of the electromagnetic current. This can be
verified by noting that
\begin{equation}
(k+k')^\mu F_\mu(k,k') \ = \ -\, i \int d^4z\; e^{i k'z}\int_0^{z(k+k')}
d\omega\; e^{-i\omega} \ = \ (2\pi)^4 \Big[\delta^{(4)}(k) -
\delta^{(4)}(k')\Big] \ ,
\label{deltas}
\end{equation}
which leads to
\begin{equation}
p^\mu \alpha_{f\,\mu}(q,p) \ = \ f(q^+) - f(q^-) \ .
\label{alphalong}
\end{equation}
It is also worth noticing that the integral in Eq.~(\ref{deltas}) becomes
trivial, therefore the result in Eq.~(\ref{alphalong}) does not depend on
the integration path in Eq.~(\ref{pathf}) [a similar mechanism leads to the
path independence of the functions in Eqs.~(\ref{f01})]. Using the relation
in Eq.~(\ref{alphalong}), after an adequate change of variables one obtains
\begin{equation}
p^\mu \Big(\Pi_{\mu\nu}^{{\rm (I)}\,a} + \Pi_{\mu\nu}^{{\rm (II)}\,a}\Big)
\ = \ 0\ .
\end{equation}
A similar cancellation has been found in Ref.~\cite{Plant:1997jr}, where a
nlNJL model that includes vector mesons without quark WFR is considered.

Let us now concentrate on the $\rho$ electromagnetic decay
constant $f_v$, which can be defined from $\rho^0\to e^+e^-$
decay:
\begin{equation}
\Gamma(\rho^0\to e^+e^-) \ = \ \frac{4\pi}{3}\,\alpha^2 \, m_\rho\,
f_v^2\ ,
\end{equation}
where $\alpha = e^2/(4\pi)$ is the electromagnetic fine structure
constant. It can be seen that $f_v$ is related to the trace of
$\Pi_{\mu\nu}^3(p)$ through
\begin{equation}
3\, m_\rho^2\, e\, f_v \ = \ g_{\mu\nu}\,\Pi_{\mu\nu}^3
(p)\Big|_{p^2 = -m_\rho^2}\ .
\end{equation}
To evaluate the transverse part of the tensor $\Pi_{\mu\nu}^3$ we take a
straight line path for the integral over $s_\mu$ in Eq.~(\ref{path}). This
leads to
\begin{equation}
\alpha_{f\,\mu}(q,p) \ = \
\int^1_{-1} d \lambda\
                       \left( q_\mu + \lambda \dfrac{p_\mu}{2}\right)
                       f^\prime\left( q + \lambda
                       \dfrac{p}{2}\right)\ ,
\end{equation}
where $f'(p)$ denotes the derivative of $f$ with respect to $p^2$.
After some algebra, we obtain
\begin{equation}
f_v \ = \ \dfrac{Z_\rho^{1/2}}{3\, m_\rho^2}
\, \left[ J^{\rm (I)} (-m_\rho^2) + J^{\rm (II)} (-m_\rho^2)\right]\ ,
\end{equation}
where
\begin{eqnarray}
J^{\rm (I)}  (p^2) &=& -\,4N_c\,\int \dfrac{d^4 q}{(2\pi)^4}\, h(q)\,\Bigg\lbrace
              \dfrac{3}{2}\,\dfrac{[z(q^+)+z(q^-)]}{D(q^+)D(q^-)}\Big[
              (q^+\cdot q^-) + m(q^+)\,m(q^-) \Big] \nonumber \\
              & & + \; \dfrac{1}{2}\,\dfrac{z(q^+)}{D(q^+)} \, +
                  \, \dfrac{1}{2}\, \dfrac{z(q^-)}{D(q^-)}\, + \,\dfrac{q^2}{(q\cdot p)}
                  \left[\dfrac{z(q^+)}{D(q^+)} - \dfrac{z(q^-)}{D(q^-)}\right]\nonumber \\
              & & + \,\dfrac{z(q^+)z(q^-)}{D(q^+)D(q^-)}\,
              \left[(q\cdot p) - \dfrac{q^2\,p^2}{(q\cdot p)}\right]\,
                  \bigg[-\,\bar\sigma_1\, \big[m(q^+) + m(q^-)\big]\,\alpha^+_g(q,p) \nonumber \\
              & & + \; \bar\sigma_2\,\big[q^2 + \dfrac{p^2}{4}
                  - m(q^+)\,m(q^-) \big]\,\alpha^+_f (q,p)\,\bigg] \Bigg\rbrace\ , \nonumber\\
J^{\rm (II)} (p^2) &=& -\,4N_c \int \dfrac{d^4 q}{(2\pi)^4}\,
              \dfrac{z(q)}{D(q)}\left\lbrace\dfrac{q^2}{(q\cdot p)}
              \Big[h(q^+)-h(q^-)\Big] + \left[(q\cdot p) -
              \dfrac{q^2\,p^2}{(q\cdot p)}\right]\alpha^+_h (q,p)
              \right\rbrace .
\end{eqnarray}
Superindices (I) and (II) correspond to the contributions from the
diagrams in Figs.~\ref{fig:ward}a and \ref{fig:ward}b,
respectively, while the functions $\alpha^+_f(q,p)$ have been
defined as
\begin{equation}
\alpha^\pm_f (q,p) \ = \ \int_{-1}^1 d\lambda \,\dfrac{\lambda}{2}\,
                   f^\prime\left( q-\lambda \dfrac{p}{2}\right) \ .
\end{equation}

\subsection{\sc $\pi^0 \rightarrow \gamma\gamma$ decay}

Let us analyze in the context of our model the anomalous decay $\pi^0\to
\gamma\gamma$. As it is well known, in the NJL model this decay is problematic:
in order to reproduce the experimentally observed result it is necessary to
perform quark loop momentum integrations up to infinity instead of following
the cutoff prescription of the model~\cite{Blin:1987hw}. In our framework,
taking into account the discussion of gauge interactions in the previous
subsections, the decay amplitude can be calculated from the matrix element
\begin{equation}
\langle 0 \vert {\cal J}_{{\rm em}\,\mu}(x) {\cal J}_{{\rm em}\,\nu}(0)
\vert \tilde \pi^3(p) \rangle \ = \frac{\delta^3 S_E^{\rm
bos}}{\delta\mathcal{A}_\mu(x)\, \delta\mathcal{A}_\nu(0)
\,\delta\tilde\pi^3(p)} \bigg\vert_{\mathcal{A}_{\mu,\nu}=\tilde\pi^3=0}\
.
\end{equation}

\begin{figure}[h]
\begin{center}
\centering
\subfloat{\includegraphics[scale=0.6]{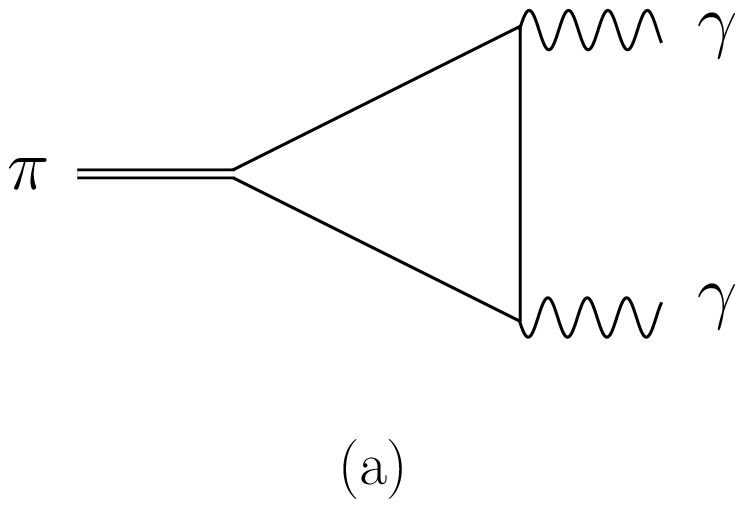}}
\subfloat{\includegraphics[scale=0.6]{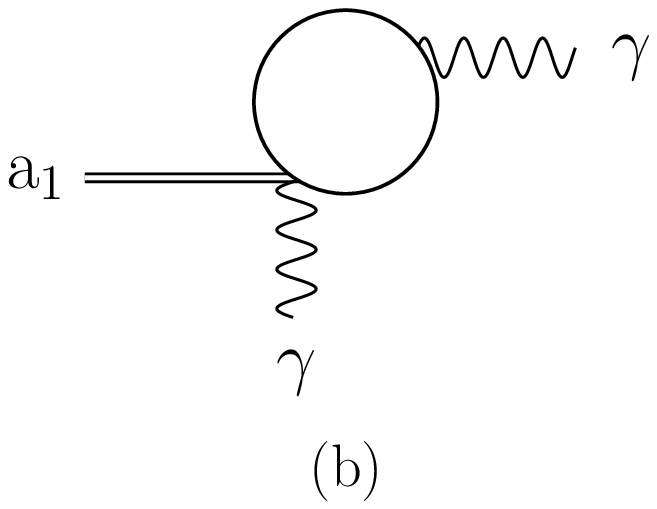}}
\subfloat{\includegraphics[scale=0.6]{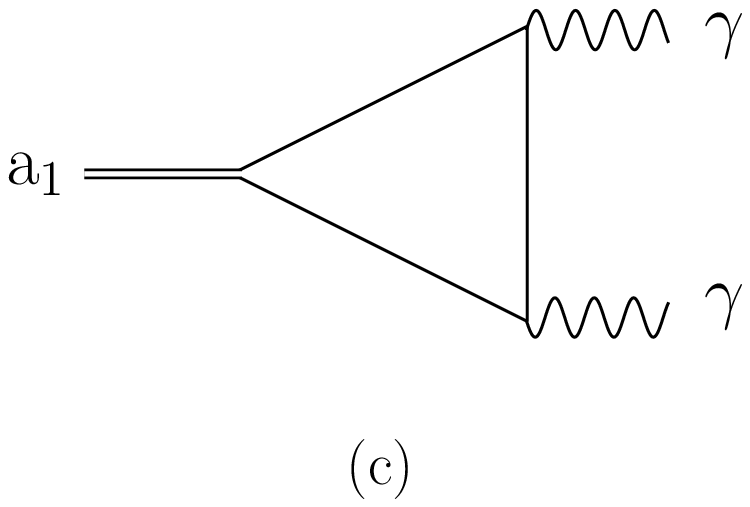}}
\end{center}
\caption{Diagrams contributing to $\pi^0 \rightarrow \gamma\gamma$ decay.}
\label{fig:pigggraf}
\end{figure}

In principle there are several diagrams that contribute to the amplitude at
the level of one loop. As in the case of the pion decay constant $f_\pi$,
since the physical $\pi^0$ state $\tilde\pi^3(p)$ is a combination of $\pi$
and $a_\mu$ fields, one has to consider the linear expansion of the
bosonized action in $\pi$ and in $a_\mu$. The diagrams leading to nonzero
contributions are those depicted in Fig.~\ref{fig:pigggraf}. If the outgoing
photons are assumed to be in states of four-momenta $k_1$ and $k_2$ with
polarization vectors $\varepsilon_\mu^{(\lambda_1)}(k_1)$ and
$\varepsilon_\nu^{(\lambda_2)}(k_2)$, respectively, the decay amplitude can
be written as
\begin{equation}
{\cal M}(\pi^0\to\gamma\gamma) \ = \ i\,4\pi\alpha\, \tilde F(k_1,k_2)\,
\epsilon^{\mu\nu\alpha\beta}\,\varepsilon_\mu^{(\lambda_1)}(k_1)^\ast
\varepsilon_\nu^{(\lambda_2)}(k_2)^\ast k_{1\alpha}\, k_{2\beta} \ ,
\end{equation}
where the form factor $\tilde F(k_1,k_2)$ is given by the sum of $\pi$ and $a_\mu$
contributions to the $\tilde\pi^3$ state,
\begin{equation}
\tilde F(k_1,k_2) \ = \ Z_\pi^{1/2}\,[F_\pi(k_1,k_2)\, +
\,\lambda(p^2)\,F_a(k_1,k_2)]\ ,
\end{equation}
with $p=k_1+k_2$.

The first term in the brackets, corresponding to the diagram in
Fig.~\ref{fig:pigggraf}a, has been calculated (apart from an isospin factor)
in Ref.~\cite{Dumm:2010hh}. One has
\begin{equation}
F_\pi(k_1,k_2) \ = \ \frac{2 N_c}{3}\,\int \dfrac{d^4
q}{(2\pi)^4}\,h\bigg(q+\frac{k_2}{2}-\frac{k_1}{2}\bigg)\,
\dfrac{z(q)z(q-k_1)z(q+k_2)}{D(q)D(q-k_1)D(q+k_2)}\,A(q,k_1,k_2)\ ,
\end{equation}
where
\begin{eqnarray}
A(q,k_1,k_2) & = & \bigg(\frac{1}{z(q)}\,+\,\frac{1}{z(q-k_1)}\bigg)
\bigg(\frac{1}{z(q)}\,+\,\frac{1}{z(q+k_2)}\bigg)\,\bigg\{m(q) \, - \, \frac{q^2}{2}
\times\nonumber\\
& & \left[
\frac{m(q+k_2)-m(q)}{(q\cdot k_2)}\,-\,\frac{m(q-k_1)-m(q)}{(q\cdot k_1)}
\right]\bigg\} \ .
\end{eqnarray}
On the other hand, the form factor $F_a(k_1,k_2)$ arises from the sum of the
contributions corresponding to the diagrams in Figs.~\ref{fig:pigggraf}b
and~\ref{fig:pigggraf}c. Although these turn out to be separately divergent,
it is seen that divergent pieces cancel out and the sum is finite. We obtain
\begin{eqnarray}
F_a(k_1,k_2) & = & -\,\frac{2 N_c}{3}\,\int \dfrac{d^4
q}{(2\pi)^4}\;\Bigg\{
h\big(q+k_2/2-k_1/2\big)\,
\dfrac{z(q)z(q-k_1)z(q+k_2)}{D(q)D(q-k_1)D(q+k_2)}\times \nonumber \\
& & \bigg[ \Big(m(q-k_1) + m(q+k_2)\Big)\,A(q,k_1,k_2) \,+ \nonumber \\
& & \frac{q^2}{2}\bigg(
\frac{B(q,q-k_1,q+k_2)}{(q\cdot k_2)}\, -\, \frac{B(q,q+k_2,q-k_1)}{(q\cdot k_1)}
\bigg)\bigg]\, +
\nonumber \\
& & q^2\, \bigg[\frac{h(q+k_2/2)}{(q\cdot k_2)}\,C(q,k_1)\, +
\,\frac{h(q+k_1/2)}{(q\cdot k_1)}\,C(q,k_2)\bigg]\Bigg\}
\ ,
\end{eqnarray}
where
\begin{eqnarray}
B(q,r,s) & = & \bigg(\frac{1}{z(q)}\,+\,\frac{1}{z(r)}\bigg)
\bigg(\frac{1}{z(q)}\,-\,\frac{1}{z(s)}\bigg)\,D(s)\ , \nonumber\\
C(q,k) & = & \bigg(\frac{1}{z(q+k/2)}\,+\,\frac{1}{z(q-k/2)}\bigg)
\,\frac{z(q+k/2)z(q-k/2)}{D(q+k/2)D(q-k/2)} \ .
\end{eqnarray}
Finally, after phase space integration and sum over outgoing photon
polarizations, the $\pi^0\to\gamma\gamma$ decay amplitude is given by
\begin{equation}
\Gamma(\pi^0\to\gamma\gamma) \ = \ \frac{\pi}{4}\,\alpha^2\,m_\pi^3\,
\tilde F(k_1,k_2)^2\ .
\end{equation}
Since photons are on-shell, from Lorentz invariance it is seen that $\tilde
F(k_1,k_2)$ can only be function of the scalar product $(k_1\cdot k_2) =
-m_\pi^2/2$.

\subsection{\sc $\rho \rightarrow \pi\pi$ decay}

In general, various transition amplitudes can be calculated by expanding the
bosonized action to higher orders in meson fluctuations. In this subsection
we concentrate in the processes $\rho^0\to\pi^+\pi^-$ and $\rho^\pm
\to\pi^\pm\pi^0$, which are responsible for more than 99\% of $\rho$ meson
decays. The decay amplitudes ${\mathcal M}(v^a(p)\to\pi^b(q_1)\pi^c(q_2))$
are obtained by calculating the corresponding functional derivatives of the
effective action, which can be written in terms of two form factors $\tilde
F_{\rho\pi\pi}(p^2,q_1^2,q_2^2)$ and $\tilde
G_{\rho\pi\pi}(p^2,q_1^2,q_2^2)$:
\begin{eqnarray}
\frac{\delta^3 S_E^{\rm bos}}{\delta\tilde v_\mu^a(p) \delta
\tilde{\pi}^b(q_1)\delta \tilde{\pi}^c(q_2)}
\bigg\vert_{\delta v_\mu=\delta\pi=0} & = & (2\pi)^4 \
\delta^{(4)} (p + q_1 + q_2) \ \epsilon_{abc} \left[\tilde
F_{\rho\pi\pi}(p^2,q_1^2,q_2^2) \; \frac{(q_{1\mu} + q_{2\mu})}{2}\right.
\nonumber \\
& & \; + \; \left.
\tilde G_{\rho\pi\pi}(p^2,q_1^2,q_2^2)\; \frac{(q_{1\mu} - q_{2\mu})}{2}\right]
\ .
\label{dsppv}
\end{eqnarray}
Only the transverse piece, driven by the form factor $\tilde
G_{\rho\pi\pi}(p^2,q_1^2,q_2^2)$, contributes to $\rho\to\pi\pi$ decay
widths. Indeed, in the isospin limit, one has
\begin{equation}
\Gamma_{\rho^0\to\pi^+\pi^-}\ = \
\Gamma_{\rho^\pm\to\pi^\pm\pi^0}\ = \ \frac{1}{48\pi}\;m_\rho\;
g_{\rho\pi\pi}^2\left(1-\frac{4m_\pi^2}{m_\rho^2}\right)^{3/2} \ ,
\end{equation}
where $g_{\rho\pi\pi} \equiv
\tilde G_{\rho\pi\pi}(-m_\rho^2,-m_\pi^2,-m_\pi^2)$.

\begin{figure}[h]
\begin{center}
\subfloat{\includegraphics[scale=0.6]{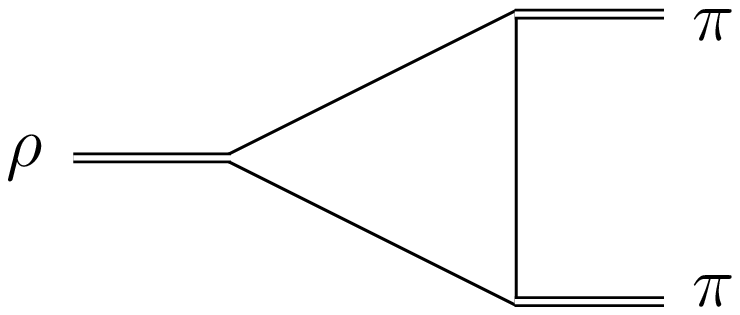}}
\subfloat{\includegraphics[scale=0.6]{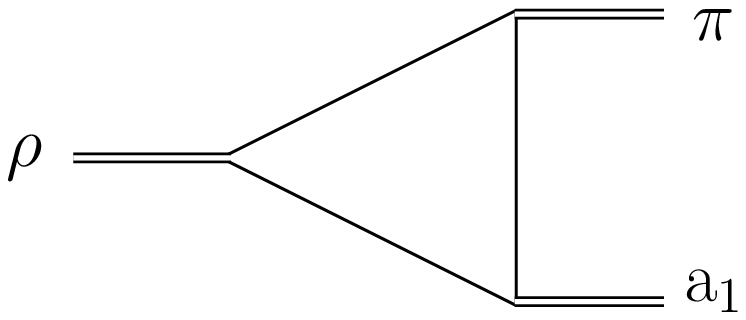}}
\subfloat{\includegraphics[scale=0.6]{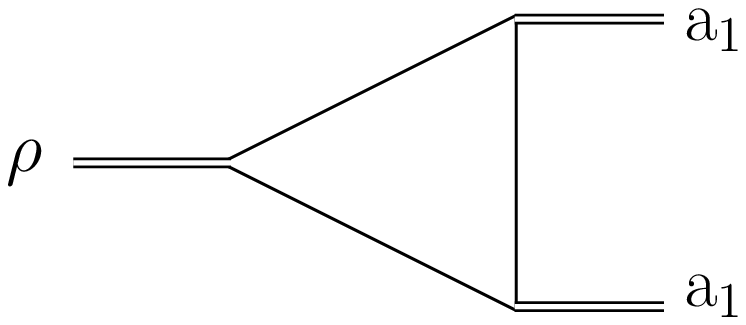}}
\end{center}
\caption{Diagrams contributing to $\rho \rightarrow \pi\pi$ decays.}
\label{fig:ropipigraf}
\end{figure}

The form factor $\tilde G_{\rho\pi\pi}(p^2,q_1^2,q_2^2)$ arises from the
effective vertex $\tilde\rho\tilde\pi\tilde\pi$, where $\tilde\rho$ and
$\tilde\pi$ are renormalized states. Since we expand the effective action
in Eq.~(\ref{sboson}) in powers of the unrenormalized fields, it is
convenient to write the effective vertex in terms of the original fields
$\rho$, $\pi$ and $a_\mu$ [the latter has to be taken into account due to
the $\pi - a$ mixing given by Eq.~(\ref{pia1mixing}), as mentioned in
previous subsections]. In this way, the form factor receives contributions
from the diagrams sketched in Fig.~\ref{fig:ropipigraf}. One has
\begin{eqnarray}
\tilde G_{\rho\pi\pi} (p^2,q_1^2,q_2^2) & = & Z_\rho^{1/2}\,Z_\pi\,
\bigg[ G_{\rho\pi\pi}(p^2,q_1^2,q_2^2) + \nonumber \\
& & \lambda(p^2) \ G_{\rho\pi a} (p^2,q_1^2, q_2^2) + \lambda(p^2)^2 \
G_{\rho a a} (p^2, q_1^2, q_2^2) \bigg] \ ,
\label{grhopipi}
\end{eqnarray}
where $G_{\rho\pi\pi}(p^2,q_1^2, q_2^2)$, $G_{\rho\pi a}(p^2,q_1^2, q_2^2)$
and $G_{\rho a a}(p^2, q_1^2, q_2^2)$ are one-loop functions that arise from
the expansion of the effective action. The explicit forms of these functions,
which can be obtained after a rather lengthy calculation, can be found in
Appendix A.

\section{\sc Numerical results}

\subsection{Model parameters and form factors}

To fully define the model it is necessary to provide the values of the
unknown parameters and to specify the shape of the form factors entering the
nonlocal fermion currents. There are six parameters, namely, the current
quark mass $m_c$ and the dimensionful coupling constants $G_S$, $G_V$,
$G_0$, $G_5$ and $\varkappa$. Regarding the form factors, as stated in the
Introduction, we will take into account the results obtained in lattice QCD
for the momentum dependence of the mass and WFR in the quark propagator.
Therefore, following Ref.~\cite{Bowman:2002bm}, we write the effective mass
$m(p)$ as
\begin{equation}
m(p) \ = \ m_c \, + \, \alpha_m\, f_m(p^2)\ ,
\label{fm}
\end{equation}
where $\alpha_m$ is a mass parameter defined by the normalization condition
$f_m(0) = 1$. Since LQCD calculations involve various current quark masses,
we have chosen to take as input the shape of the (normalized) function
$f_m(p^2)$, taking LQCD results in the limit of low $m_c$ and smallest lattice
spacing. Considering the LQCD analysis in Ref.~\cite{Bowman:2002bm}, we
parameterize this function by
\begin{equation}
f_m(p^2) \ = \ \frac{1}{1 + (p^2/\Lambda_0^2)^\alpha}\ ,
\label{fmp}
\end{equation}
with $\alpha = 3/2$.
On the other hand, for the wave function renormalization we use
the parametrization~\cite{Noguera:2005ej,Noguera:2008}
\begin{equation}
z(p) \ = \ 1 \, - \, \alpha_z\, f_z(p^2)\ ,
\label{fz}
\end{equation}
where
\begin{equation}
f_z(p^2) \ = \ \frac{1}{\left(1 + p^2/\Lambda_1^2\right)^{\beta}}\ .
\label{fzp}
\end{equation}
It is found that LQCD results favor a relatively low value for the exponent
$\beta$, therefore we take here $\beta = 5/2$, which is the smallest
exponent compatible with the ultraviolet convergence of the gap equations
(\ref{gapeq}). As required by dimensional analysis and Lorentz invariance,
the functions $f_m(p^2)$ and $f_z(p^2)$ carry dimensionful parameters
$\Lambda_0$ and $\Lambda_1$, which represent effective cutoff momenta in the
corresponding channels. Thus, we will use here the above functional forms
for the form factors, taking $\Lambda_0$ and $\Lambda_1$ as two further free
parameters of the model. Regarding the parameters $\alpha_m$ and $\alpha_z$
introduced in Eqs.~(\ref{fm}) and (\ref{fz}), from Eqs.~(\ref{mz}) it is
seen that they are related to the mean field values of the scalar fields by
\begin{eqnarray}
m(0) & = & m_c + \alpha_m \ = \ \frac{m_c + \bar \sigma_1}{1 -
\bar\sigma_2} \ , \\
z(0) & = & 1 - \alpha_z \ = \ \frac{1}{1 - \bar\sigma_2} \ ,
\label{amaz}
\end{eqnarray}
hence, for a given set of model parameters, they can be obtained by solving
the gap equations~(\ref{gapeq}).

The model also includes the form factors $h(p)$, $h_0(p)$ and $h_5(p)$,
introduced through the vector and axial vector current-current interactions.
For definiteness and simplicity we will assume the effective behavior of
quark interactions to be similar in the $J=0$ and $J=1$ channels, therefore
we will take for $h(p)$ the same form as $g(p)$. Regarding the
vector-isoscalar sector, as it is usually done we assume approximate
degeneracy with the vector-isovector part, hence we take $h(p)\simeq
h_0(p)$. The axial vector-isoscalar sector can be studied separately, since
it decouples from the rest of the Lagrangian. Here we will just take $h_5(p)
= h(p)$ in order to get an estimation for the constant $G_5$ from
phenomenology.

\begin{figure}[h]
\begin{center}
\includegraphics[width=1.\textwidth]{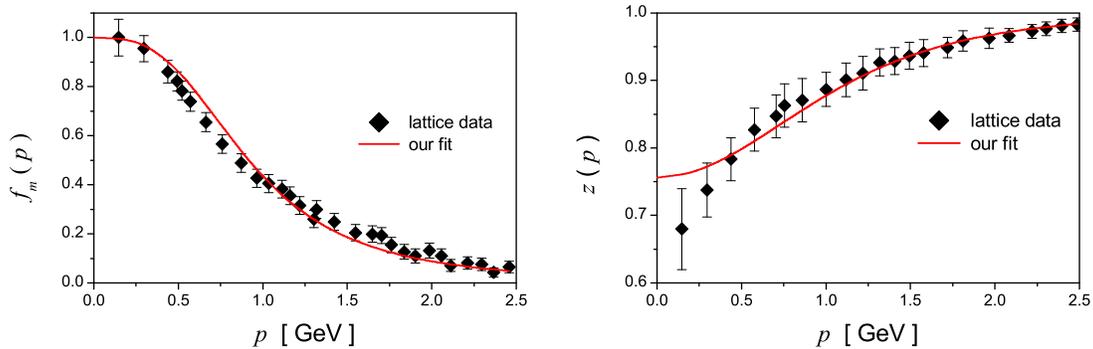}
\end{center}
\vspace{-0.9cm}
\caption{Fit to lattice data for the functions $f_m(p^2)$ and
$z(p)$. }
\label{fig:lattice}
\end{figure}

Given the form factor shapes, in order to study the phenomenology we have to
determine the values of the model parameters (current quark mass, coupling
constants and effective cutoff momenta). To do this, we first carry out a
fit to lattice results for the functions $f_m(p^2)$ and $z(p)$, from which
we obtain the values of the cutoffs $\Lambda_0$ and $\Lambda_1$, as well as
the parameter $\alpha_z$. The latter will be used, together with five
phenomenological quantities, as input to determine the remaining six free
model parameters. From the LQCD results quoted in
Ref.~\cite{Parappilly:2005ei} we obtain
\begin{equation}
\Lambda_0 = 917\,\pm\, 14 \, {\rm \ MeV}\ , \qquad
\Lambda_1 = 1775\,\pm\, 53 \, {\rm \ MeV}\ , \qquad
\alpha_z = 0.244\,\pm\, 0.010 \, \ ,
\label{fit}
\end{equation}
with $\chi^2/{\rm dof} = 1.17$ and $\chi^2/{\rm dof} = 0.25$ for the fits to
$f_m(p^2)$ and $z(p)$ data, respectively. The fits have been carried out
considering lattice values up to 2.5 GeV. Both the data and the fitting
curves for $f_m(p^2)$ and $z(p)$ are shown in Fig.~\ref{fig:lattice}. In the
case of $z(p)$, it is seen that the fit leads to somewhat large values of
$z(p)$ at low momenta in comparison with lattice points. We notice, however,
that errors in this region are relatively large, and in addition these
points are the most sensitive to changes in lattice spacing and/or sea quark
masses~\cite{Parappilly:2005ei}.

Once the form factor shapes have been fixed, one can set the model
parameters so as to reproduce the empirical values of some selected
observables. As stated, we take from the fit the values of $\Lambda_0$ and
$\Lambda_1$ and then we determine the values of the parameters $m_c$, $G_V$,
$G_S$, $G_0$, $G_5$ and $\varkappa$ from six input quantities. These have
been chosen to be the fitted value of $\alpha_z$ together with the empirical
values of the pion weak decay constant $f_\pi$ and the masses of the $\pi$,
$\rho$, $\omega$ and $f_1$ mesons. From our numerical analysis we find that
there is a set of parameters that allows us to properly reproduce these
empirical values. The corresponding results are quoted in Table~I.
\begin{table} [h]
\begin{center}
\begin{tabular}{c c | c c c c}
\hline
\ \ Model parameters\ \ & \ \ Inputs\ \ &
\multicolumn{2}{c}{\ \ \ \ Model parameters\ \ \ \ } & \multicolumn{2}{c}{Inputs} \\
\hline
$\Lambda_0$ [MeV] & \ \ LQCD results \ \ & \ \ $m_c$ [MeV] \ \ & 1.59 & $\alpha_z$ & LQCD results \\
$\Lambda_1$ [MeV] & \ \ LQCD results \ \ & $G_S\Lambda_0^2$ & 19.0 & \ \ $m_\pi$ [MeV] \ \ & 139 \\
& & $\varkappa/\Lambda_0$ & 11.2 & $f_\pi$ [MeV] & 92.2 \\
& & $G_V\Lambda_0^2$ & 13.0 & $m_\rho$ [MeV] & 775 \\
& & $G_0\Lambda_0^2$ & 12.8 & $m_{f_1}$ [MeV] & 1280 \\
& & $G_5\Lambda_0^2$ & $\sim 14$ & $m_\omega$ [MeV] & 783 \\
\hline
\end{tabular}
\caption{Model parameters. The values of $\Lambda_0$, $\Lambda_1$ and
$\alpha_z$ have been obtained from a fit to lattice QCD calculations for the
effective quark propagator, see Eq.~(\ref{fit}). The model parameters $m_c$,
$G_S$, $\varkappa$, $G_V$, $G_0$ and $G_5$ are fitted against the
phenomenological values of five hadronic observables, plus the value of
$\alpha_z$ given by the fit to LQCD data.}
\end{center}
\label{tab:param}
\end{table}

The numerical analysis requires solving a system of coupled equations that
includes the gap Eqs.~(\ref{gapeq}), equations $G_{M}(-m_M^2) = 0$ for
$M=\pi,\rho,\omega$ and $f_1$ to determine meson masses, and Eq.~(\ref{fpi})
for $f_\pi$. This involves the calculation of one-loop integrals introduced
in Secs.~III.A and III.B, which in general is not a trivial task due to the
fact that the form factor $f_m(p^2)$, as function of the fourth component
$p_4$ of the momentum, has cuts when $p_4$ is extended to the complex plane.
Depending on the value of the three-momentum $\vec p$ these cuts can
occasionally cross the real axis, and have to be taken into account through
a proper deformation of the integration path. Details of the calculations
are given in Appendix~B.

From Table~I we find a ratio $G_S/G_V \sim 1.5$, which is in agreement with
standard NJL model parametrizations~\cite{njlrev}. Concerning the value of
$G_0$, it is necessary to take into account that we are working within a
two-flavor model, therefore effects of strange quark bound states are not
explicitly considered. Our determination of $G_0$ would be valid only in the
case of ``ideal mixing'' between SU(3)$_f$ singlet and octet $I=0$ states,
which means taking the $\omega$ as an approximate SU(3)$_f$ octet state, and
the $\phi$ meson as an approximately pure $\bar s s$ state. In the case of
the $f_1$ axial vector meson there is an additional problem, which is common
to various quark models. Indeed, models that do not include an explicit
mechanism of confinement usually have difficulties for describing meson
resonances, since there is a threshold above which the meson mass becomes
large enough to allow the decay of the meson into two quarks. This threshold
is typically of the order of $2m(0)$, therefore models that lead to
constituent masses larger than about 400 MeV (as occurs in our case) can
avoid this problem for low mass resonances like the $\rho$
meson~\cite{He:1997gn}. Other possible approaches are e.g.~the extension of
$G_M(-s)$ functions to the complex plane~\cite{Zhuang:1994dw} or the search
for a peak in the meson spectral function~\cite{Hansen:2006ee}.
Mathematically, in our model the onset of the unphysical $q\bar q$ channel
corresponds to the fact that in the integrals of the form of
e.g.~Eq.~(\ref{grho}) there is a ``pinch point'' at which both functions
$D(q^+)$ and $D(q^-)$ in the integrand are equal to zero (i.e.~both
constituent quarks are simultaneously on shell). For the parameters in Table
I, the threshold is found to be at 1264~MeV, i.e.~below the empirical value
$m_{f_1}= 1280$ MeV, and the free parameter to be adjusted to get the
phenomenological value of the $f_1$ mass is the coupling constant $G_5$. To
obtain an approximate value for this constant, we have solved the equation
$G_{f_1}(-m^2)=0$ varying $G_5\Lambda_0^2$ from large values of $G_5$ up to
$G_5\Lambda_0^2\simeq 22$, which leads to $m\simeq 1$ GeV, and then we have
extrapolated to the region above the threshold to obtain $m_{f_1}\simeq
1280$~MeV for $G_5\Lambda_0^2\sim 14$.

\subsection{Numerical results for phenomenological quantities}

Using the parameters and nonlocal form factors quoted in the previous
subsection, we can calculate the predictions of the model for the
phenomenological quantities analyzed in Secs.~II and III.

Our numerical results for various observables are summarized in Table~II (we
have not included here the quantities taken as phenomenological inputs,
namely $m_\pi$, $f_\pi$, $m_\rho$, $m_\omega$ and $m_{f_1}$). From the table
it is seen that the predictions of the model for the $\pi^0\to\gamma\gamma$,
$\rho\to e^+e^-$ and $\rho\to\pi\pi$ decay rates are in good agreement with
experiments, being compatible with the empirical values~\cite{pdg} within an
accuracy of less than 10\%. We can also obtain a prediction for the width
$\Gamma(\omega\to e^+e^-)$, which is found to be about 0.8~keV, somewhat
larger than the experimental value $0.60\pm 0.02$~keV~\cite{pdg}. However,
as discussed above, our result might become modified after the inclusion of
strangeness degrees of freedom owing to the $\omega-\phi$ mixing. Regarding
the $\sigma-\sigma'$ sector, we obtain a physical state with a mass of about
680 MeV, which can be identified with the observed $\sigma$ meson resonance
(the mass of which is rather uncertain), while for the state $\sigma'$ we
find that the function $G_{\sigma'}(-s)$ grows monotonically with $s$,
indicating that this state does not represent a physical meson (a more
detailed discussion on the $\sigma'$ state in this type of models can be
found in Ref.~\cite{Noguera:2008}). In the case of the a$_1$ vector mesons
we find that the function $G_{{\rm a}_1} (-s)$ decreases with $s$ until it
reaches a minimum at $\sqrt{s}\simeq 1250$~MeV, very close to the threshold
of on-shell quark pair production, or pinch point, found at 1264~MeV.
Recalling the discussion in the previous subsection, in order to estimate
the value of the ${{\rm a}_1}$ mass it is possible either to take the
minimum of $G_{{\rm a}_1} (-s)$ or to make an extrapolation based on the
behavior of $G_{{\rm a}_1} (-s)$ up to say $s\sim (1\ {\rm GeV})^2$. Both
approaches lead to $m_{{\rm a}_1}\sim 1200-1250$ MeV, which is in good
agreement with experimental expectations. We have also analyzed the
dependence of our results on the value of $\alpha_z$ within the error given
by the fit to LQCD data [see Eq.~(\ref{fit})], obtaining that the model
predictions do not vary significantly.

\begin{table} [H]
\begin{center}
\begin{tabular}{cc|c|c}
\hline
\hspace{-2cm} & \hspace{5.2cm} & \hspace{.8cm} Model \hspace{.8cm} & \hspace{1cm} Empirical \hspace{1cm} \\
\hline
\hspace{-2cm} &$\Gamma(\pi^0\to \gamma\gamma)$ [MeV] \hspace{-1.2cm} & $7.82\times 10^{-6}$ & $(7.63\pm 0.16)\times 10^{-6}$ \\
\hspace{-2cm} &$\Gamma(\rho\to e^+ e^-)$ [MeV] \hspace{-1.2cm} & $6.71\times 10^{-3}$ & $(7.04\pm 0.06)\times 10^{-3}$ \\
\hspace{-2cm} &$\Gamma(\rho\to\pi\pi)$ [MeV] \hspace{-1.2cm} & 137 & $149.1\pm 0.8$ \\
\hspace{-2cm} &$m_\sigma$ [MeV] \hspace{-1.2cm} & 683 & 400$\,$-$\,$550 \\
\hspace{-2cm} &$m_{{\rm a}_1}$ [MeV] \hspace{-1.2cm}  & 1200$\,$-$\,$1250 \ \ \ & 1190$\,$-$\,$1270 \\
\end{tabular}
\caption{\small{Model predictions and empirical values~\cite{pdg} for various observables.}}
\end{center}
\label{tab:prop}
\end{table}

Finally, in Table~III we quote our results for mean field values of scalar
fields, chiral quark condensates and effective quark-meson couplings. It is
seen that the model leads to a zero-momentum effective quark mass $m(0) =
(m_c + \bar\sigma_1)/(1-\bar\sigma_2) \simeq 400$ MeV, somewhat larger than
the value of 311~MeV obtained in Ref.~\cite{Noguera:2008} for a nlNJL model
without vector meson degrees of freedom. For comparison, notice that
standard NJL model parametrizations lead to values of constituent
(momentum-independent) quark masses around 350~MeV~\cite{njlrev}. Concerning
the chiral quark condensates, our results are relatively large in comparison
with usual phenomenological estimations and lattice calculations, which lead
to condensates in the range of $(-240\ \rm{MeV})^3$ to $(-320\
\rm{MeV})^3$~\cite{McNeile:2005pd}. In addition, when determining the model
parameters we have found a relatively low value for the current quark mass,
namely $m_c = 1.59$~MeV, in comparison with lattice estimates that lead to
$m_c\simeq 3.4\pm 0.25$~MeV in the isospin limit~\cite{pdg}. The results for
these quantities in nlNJL models are in fact strongly dependent on the form
factor shapes, as it is found in
Refs.~\cite{Noguera:2008,Carlomagno:2013ona,Hell:2011ic}, where two- and
three-flavor nonlocal models (which do not include the vector meson sector)
are considered. As discussed in those articles, one has to take into account
that both $m_c$ and $\langle \bar q q\rangle$ are scale-dependent
quantities, and our fit has been carried out using lattice data that
correspond to a renormalization scale $\mu = 3$~GeV, somewhat larger than
the usual scale of 2~GeV. To get rid of the scale dependence one can look at
the product $-\langle\bar q q\rangle m_c$, for which we get, within our
parametrization, a result of about $8.12\times 10^{-5}$ GeV$^4$. This is in
good agreement with the value arising from the Gell-Mann-Oakes-Renner
relation at the leading order in the chiral expansion, namely $-\langle\bar
q q\rangle m_c = f^2_\pi m_\pi^2/2 \simeq 8.21 \times 10^{-5}$ GeV$^4$.
Finally, for completeness we include in Table III the values obtained for
the effective quark-meson couplings $g_{\pi q\bar q}$ and $g_{\rho q\bar
q}$.

\begin{table} [H]
\begin{center}
\begin{tabular}{cc|c}
\hline
\hspace{-2cm} & \hspace{5.2cm} & \hspace{.8cm} Model \hspace{.8cm} \\
\hline
\hspace{-2cm} & $\bar{\sigma}_1$ [MeV] \hspace{-1.2cm} & 524 \\
\hspace{-2cm} & $\bar{\sigma}_2$ & -0.322 \\
\hspace{-2cm} & $-\langle \bar q q \rangle ^{1/3}$ [MeV] \hspace{-1.4cm} & 371 \\
\hspace{-2cm} & $g_{\pi q\bar{q}}$ & 5.69 \\
\hspace{-2cm} & $g_{\rho q\bar{q}}$ & 2.94 \\
\end{tabular}
\caption{\small{Numerical results for various phenomenological quantities.}}
\end{center}
\label{tab:cond}
\end{table}

\section{\sc Summary \& outlook}

In this work we have introduced a two-flavor chiral quark model that
includes nonlocal four-fermion interactions. Besides the usual scalar and
pseudoscalar couplings already present in the standard (local) NJL model, we
consider the couplings between vector and axial-vector quark-antiquark
currents as well as a current-current interaction that leads to WFR of the
quarks fields. The model leads to a dressed quark propagator in which the
effective mass and WFR are functions of the momentum through nonlocal form
factors, and these can be fitted to the results obtained in lattice QCD
calculations.

We have concentrated on vacuum properties related with the presence of
vector and axial-vector mesons, which have not been taken into account in
this context in previous works. For this analysis we have evaluated various
one-loop diagrams contributing to vector and axial-vector mass terms and
decay amplitudes. It is seen that, owing to the nonlocal character of the
interactions, the model leads to tadpole diagrams contributing to the
$\rho-$photon vertex, in addition to the usual quark loop contributions. The
longitudinal components of both contributions are found to be separately
nonvanishing, while their sum is transverse, as requested by electromagnetic
current conservation. It is worth mentioning that analytical expressions for
the pion mass and decay constants obtained in previous works have been
revisited in order to take into account $\pi-$a$_1$ mixing.

On the phenomenological side, the fit of nonlocal form factors to lattice
QCD results for effective quark propagators provides a more natural and
realistic way to regularize the model in comparison with the standard NJL
approach. The remaining unknown parameters, namely the current quark mass
and the current-current coupling constants, can be determined from some
input observables. Here we have chosen to take as inputs the measured values
of the pion decay constant and a set of meson masses. From the numerical
evaluation of the analytical expressions we find that the model is able to
properly reproduce the empirical values of these observables, and leads to
phenomenologically acceptable values for other scalar and vector meson
masses and decay widths.

To conclude, let us state that the inclusion of the axial and vector meson
sector offers a more complete picture of hadron phenomenology in the
framework of nonlocal quark models, and its effects can be important for the
analysis of hadronic observables such as the pion electromagnetic form
factor and the vector and axial vector form factors for pion radiative
decays. It is also worth it to extend the study of $\rho$ meson properties to
finite-temperature systems, given its importance for the study of heavy ion
collisions. In addition, for the case of hadronic systems at finite chemical
potential it is expected that vector interactions lead to a nonzero
condensate in the $J=1$, $I=0$ channel, which can be important for the study
of the QCD phase diagram~\cite{Hell:2012da} and the physics of compact
objects~\cite{Blaschke:2007ri}. We expect to report on these issues in
forthcoming articles.

\section*{Acknowledgments}

We are grateful to S. Noguera for valuable comments and discussions. This
work has been partially funded by CONICET (Argentina) under Grants No.\ PIP
578 and PIP 449, by ANPCyT (Argentina) under Grants No.\ PICT-2011-0113 and
PICT-2014-0492, and by the National University of La Plata (Argentina),
Project No.\ X718.

\section*{\sc Appendix A: Analytical expressions for the form factors in $\rho\to\pi\pi$ decays}

Here we quote the analytical expressions for the functions
$G_{\rho\pi\pi}(p^2,q_1^2,q_2^2)$, $G_{\rho\pi a}(p^2,q_1^2,q_2^2)$ and
$G_{\rho aa}(p^2,q_1^2,q_2^2)$ contributing to the form factor $\tilde
G_{\rho\pi\pi}(p^2,q_1^2,q_2^2)$, see Eq.~(\ref{grhopipi}). To calculate the
$\rho\to\pi\pi$ decay amplitude, we have to evaluate these functions at
$q_1^2 = q_2^2 = (p-q_1)^2 = -m_\pi^2$, $p^2 = -m_\rho^2$. We find it
convenient to introduce the momentum $v = q_1 - p/2$, which satisfies
$p\cdot v = 0$, $v^2 = m_\rho^2/4 -m_\pi^2$. Then the functions $G_{\rho x
y}(p^2,q_1^2,q_2^2)$, where subindices $x$ and $y$ stand for either $\pi$ or
$a$, can be written as
\begin{eqnarray}
G_{\rho x y}(p^2,q_1^2,q_2^2) & = & 16N_c \, \int \dfrac{d^4 q}{(2\pi)^4}
\,h(q)\,g\left(q + v/2 + p/4\right)\,g\left(q + v/2 -
p/4\right) \nonumber \\
& & \times\ \dfrac{z(q^+)z(q^-)z(q+v)}{D(q^+)D(q^-)D(q+v)}
\; f_{x y}(q,p,v) \ ,
\label{grhopipiint}
\end{eqnarray}
where we have defined $q^\pm = q \pm p/2$. After a rather lengthy
calculation we find for $f_{xy}(q,p,v)$ the expressions
\begin{eqnarray}
f_{\pi\pi} &=& \bigg[(q^+ \cdot q^-) + m(q^+)\,m(q^-)\bigg] \, \bigg[1 + \dfrac{(q\cdot v)}{v^2}\bigg]
\nonumber \\
& & -\, \dfrac{(q\cdot v)}{v^2} \bigg\{ 2\, \Big[\,q\cdot
(q+v)\Big] \, + \, m(q+v)\,\Big[m(q^+)+\,m(q^-)\Big]\bigg\} \ ,\nonumber \\
f_{\pi a} &=& - 2\, m(q+v) \left[ (q^+\cdot q^-)\, - \,2\, \dfrac{(q\cdot v)^2}{v^2}
\, + \, m(q^+)m(q^-)\right] \nonumber \\
& & + \, \bigg[1 + \dfrac{(q\cdot v)}{v^2}\bigg] \, \bigg\{(q^+\cdot p)\,m(q^-)
- (q^-\cdot p)\,m(q^+)\,-\,2(q \cdot v) \Big[m(q^+)+m(q^-)\Big]\bigg\} \ ,
\nonumber \\
f_{aa} &=&
\bigg[1 + \dfrac{(q\cdot v)}{v^2} \bigg] \bigg[q^{+2}\,q^{-2}\, - \, (q^+\cdot
q^-) \, (q+v)^2 \, - \, \Big( v^2 + \dfrac{p^2}{4} \Big) m(q^+)m(q^-) \bigg]
\nonumber \\
& & + \; m(q+v) \bigg\{ m(q^+)\, (q^- \cdot p)\, -\, m(q^-)\, (q^+ \cdot p)\,
+\, \dfrac{(q\cdot v)}{v^2}\bigg(v^2 - \dfrac{p^2}{4} \bigg)\,
\Big[ m(q^+)\, + \, m(q^-)\Big] \bigg\} \nonumber \\
& & + \; 2\,\dfrac{(q\cdot v)}{v^2} \,(q+v)^2 \bigg[(q\cdot v) - \dfrac{p^2}{4}
\bigg]\ .
\end{eqnarray}

\section*{\sc Appendix B: Loop integrals and branch cuts in the form factors}

As described in Sec.~IV, we have considered a parametrization of the nlNJL
model that allows us to reproduce LQCD results for the momentum dependence of
effective quark propagators. From the comparison with LQCD data, the form
factors $g(p)$ and $f(p)$ have been written in terms of the functions
$f_m(p^2)$ and $f_z(p^2)$ given by Eqs.~(\ref{fmp}) and (\ref{fzp}). In this
appendix we discuss the numerical evaluation of loop integrals, which have
to be treated with some care given the particular form of $f_m(p^2)$.

Let us consider loop integrals that involve an external momentum $p$, such
as those in the functions $G_M(p^2)$, $F_{0,1}(p^2)$ and $J^{({\rm I},{\rm
II})}(p^2)$, defined in Sec.~III. The integrals can be generically written
as
\begin{equation}
I(p^2)\ = \ \int \dfrac{d^4 q}{(2\pi)^4}\; F(q^+, q^-,p) \ ,
\end{equation}
where $q^{\pm}=q \pm p/2$, and $F(q^+, q^-,p)$ is a function that includes
the form factors either explicitly or through the quark effective masses
and/or wave function renormalizations. More precisely, it is seen that in
general $F(q^+, q^-,p)$ may include the form factors $f_m(s)$ evaluated at
$s=(q^+)^2$, $(q^-)^2$ and/or $q^2$. We are interested in this form factor
since its explicit form $f_m(s) = 1/[1 + (s/\Lambda_0^2)^{3/2}]$ implies the
existence of a branch cut in the complex plane $s$, namely at Re$(s)<0$,
Im$(s)=0$. It is worth noticing that in all cases the integrals have to be
evaluated numerically at $p^2 = -M^2$, where $M$ is some meson mass.

To perform the calculations we choose, as usual, the 4th axis in the
direction of the external momentum. Thus one has $p^\mu = (i M,\vec 0)$, and
$I(p^2)$ can be reduced to a double integral in $q_4$ and $|\vec q|$. Since
the functions $F(q^+, q^-,p)$ are symmetric under the exchange
$q^+\leftrightarrow q^-$, it is easy to see that $F(q^+,
q^-,p)=F({q^+}^\ast,{q^-}^\ast,p)$, which ensures the reality of $I(q^2)$.
Now let us take $|\vec q|$ fixed, and consider the analytical structure of
the integrand in the complex $q_4$ plane. It is immediately seen that we
will find a pair of branch cuts in this plane arising from the function
$f_m(q^2)$, and other pairs of cuts will appear from the occurrences of
$f_m[(q^+)^2]$ and $f_m[(q^-)^2]$, respectively. In the case of $f_m(q^2) =
f_m(q_4^2+|\vec q|^2)$, the cuts are given by Re$(q_4)=0$, $|{\rm
Im}(q_4)|>|\vec q|$, hence they never cross the real $q_4$ axis, along which
the integral is to be performed. On the other hand, for $f_m[(q^\pm)^2]$ the
cuts are located at Re$(q_4)=0$, $|{\rm Im}(q_4)\pm M/2|>|\vec q|$,
therefore if $|\vec q|<M/2$, both $f_m[(q^+)^2]$ and $f_m[(q^-)^2]$ have
cuts that cross the real $q_4$ axis.

\begin{figure}[h]
\hspace*{0.7cm}
\subfloat{\includegraphics[scale=0.57]{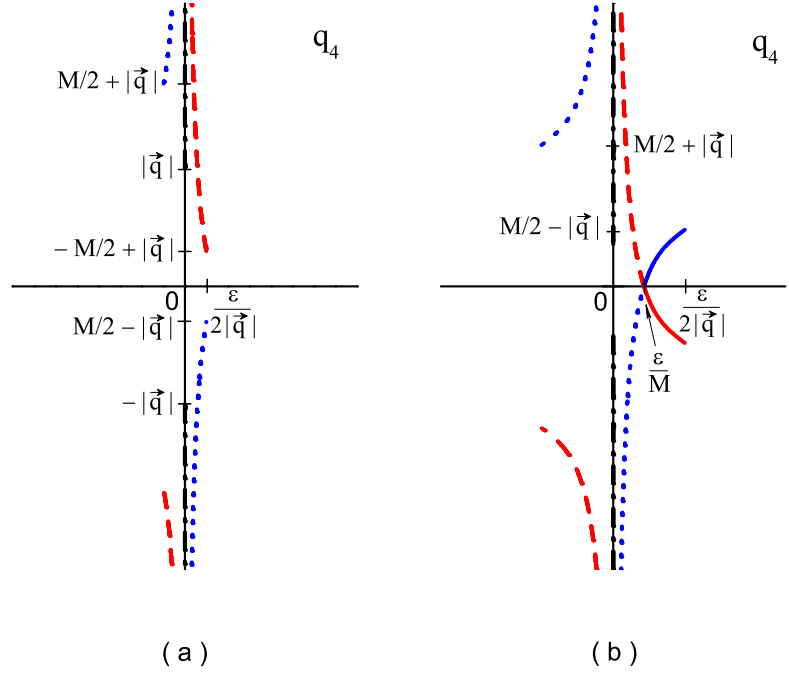}}
\vspace{-.1cm}
\caption{Branch cuts of the functions $F(q^+, q^-,p)$ in the complex plane
$q_4$, according to the prescription in Eqs.~(\ref{cutmas}) and
(\ref{cutmen}). The curves in graphs (a) and (b) correspond to $|\vec
q|>M/2$ and $|\vec q|<M/2$, respectively.} \label{fig:cortes}
\end{figure}

The treatment of these cuts is a matter of prescription. In fact, after
taking the form factors from LQCD calculations in Euclidean space, one could
turn back to Minkowski space through a Wick rotation. Then one would find
that the cuts are located along the integration axis, and to evaluate the
integrals they have to be moved away according to some recipe. Here we will
adopt the prescription of translating the arguments of $f_m(s)$ according to
\begin{eqnarray}
f_m[(q^+)^2]  & \rightarrow & f_m[(q^+)^2-i\varepsilon] \ ,
\label{cutmas}\\
f_m[(q^-)^2]  & \rightarrow & f_m[(q^-)^2+i\varepsilon] \ ,
\label{cutmen}
\end{eqnarray}
while $f_m(q^2)$ is kept unchanged. In this way, branch cuts do not overlap
and the property $F(q^+, q^-,p)=F({q^+}^\ast,{q^-}^\ast,p)$ remains valid.
From Eqs.~(\ref{cutmas}) and (\ref{cutmen}) the cuts associated to the
functions $f_m[(q^\pm)^2]$ are given by
\begin{equation}
\left\{\begin{array}{rcl}
{\rm Re}(q_4) - \dfrac{\varepsilon}{M\pm 2\,{\rm Im}(q_4)} & = & 0 \ \ ,\\
\hspace{0.5cm}\rule{0cm}{0.7cm}|{\rm Im}(q_4)\pm M/2| - |\vec q| & > & 0 \ \
.
\end{array} \right.
\label{cuts}
\end{equation}
The corresponding curves in the complex plane $q_4$ are sketched in
Fig.~\ref{fig:cortes}, where we have distinguished two situations in which
$|\vec q|>M/2$ (Fig.~\ref{fig:cortes}a) and $|\vec q|<M/2$
(Fig.~\ref{fig:cortes}b). Branch cuts corresponding to the functions
$f_m[(q^+)^2]$, $f_m[(q^-)^2]$ and $f_m(q^2)$ have been represented with
dashed, dotted and dashed-dotted lines, respectively. If $|\vec q|>M/2$, as
it is shown in Fig.~\ref{fig:cortes}a, the cuts do not cross the integration
axis, thus there is no extra contribution to the loop integral. On the
contrary, for $|\vec q|<M/2$ two branch cuts cross from one half-plane to
the other one, passing through the real $q_4$ axis. Since the integral over
$q_4$ has to be ultimately equivalent to an integral over the Minkowski
momentum $q_0$, obtained through the corresponding Wick rotation, the
integration contour along $q_4$ should be deformed in order to subtract the
contribution of the crossing pieces, which are represented with solid lines
in Fig.~\ref{fig:cortes}b. A similar procedure has to be followed when poles
of the integrand cross the integration axis at some value of $|\vec q|$; in
that case the contributions resulting from the deformation of the $q_4$
integration contour can be obtained by calculating the residues of the
poles, according to Cauchy's theorem. The need to add cut or pole
contributions to the loop integrals becomes evident by looking at relatively
simple integrals as those appearing in the gap equations (\ref{gapeq}): if
one carries out a translation of the loop momentum $p\to p\,'=p+r$, with
$r^2 = -M^2$, for fixed $|\vec p\,'|$ there will be branch cuts in the
complex plane $p\,'_4$ that cross from the upper half-plane to the lower one
(or vice versa). In addition, in general the integrand will have poles that
for large enough values of $M$ cross the real $p\,'_4$ axis at some value of
$|\vec p\,'|$. From Cauchy's theorem it is easy to see that the
corresponding contributions have to be subtracted if one requires the loop
integral to be invariant under the translation.

In practice the contributions from the cuts can be obtained by carrying out
integrations in the $q_4$ plane along adequate contours that enclose the
crossing pieces, letting then $\varepsilon\to 0$. Owing to the symmetry of
the functions $F(q^+, q^-,p)$ imaginary parts from the integrations in the
upper and lower half-planes cancel out, leading to a real total
contribution. Then the result has to be integrated over the three-momentum
variable $|\vec q|$. Notice that ---according to the conditions in
Eq.~(\ref{cuts})--- this integration goes from $|\vec q|=0$ to $|\vec
q|=M/2$, therefore the contribution can be neglected if the meson mass $M$
is relatively small, which is in general the case when $M=m_\pi$. Finally,
in the case of the $\rho\to\pi\pi$ form factor the situation is more
complicated since the relevant loop integral, given by
Eq.~(\ref{grhopipiint}), involves two independent external momenta $p$ and
$v$. It can be seen that the integrand has two additional branch cuts in the
$q_4$ complex plane, arising from the functions $f_m(s)$ evaluated at
$s=(q+v/2\pm p/4)^2$. To deal with these new cuts we have used the
prescription $f_m[(q+v/2\pm p/4)^2]  \rightarrow f_m[(q+v/2\pm p/4)^2\pm
i\varepsilon']$, choosing an integration path that encloses the pieces of
the cuts that cross the real $p_4$ axis as explained above.

\end{document}